\documentclass[12pt]{article}
\usepackage{latexsym}
\newcommand{\be}{\begin{equation}}

\newcommand{\BM}{Bohmian mechanics}

\renewcommand{\c}{classical}

\newcommand{\cb}{classical behavior}
\newcommand{\com}{center of mass}
\newcommand{\conj}{conjecture}

\newcommand{\cm}{classical motion}
\newcommand{\CM}{classical mechanics}
\newcommand{\cl}{classical limit}

\renewcommand{\d}{\delta}
\newcommand{\D}{\Delta}
\newcommand{\db}{de Broglie}
\newcommand{\deco}{decoherence}
\newcommand{\dfc}{deviation {}from classicality}

\newcommand{\e}{\epsilon}
\newcommand{\ee}{\end{equation}}

\newcommand{\env}{environment}

\newcommand{\ex}{\textrm{e}}

\newcommand{\extp}{external potential}

\renewcommand{\Im}{\mbox{Im}}
\newcommand{\iwf}{initial wave function}
\newcommand{\h}{\hbar}

\newcommand{\HJ}{Hamilton-Jacobi}

\renewcommand{\l}{\lambda}
\newcommand{\la}{\langle}

\newcommand{\LPW}{local plane wave}
\newcommand{\LPWs}{local plane waves}

\newcommand{\ode}[2]{\frac{ d #1}{ d #2}}
\newcommand{\pd}{probability distribution}
\newcommand{\pde}[2]{\frac{\partial #1}{\partial #2}}

\newcommand{\q}{quantum}
\newcommand{\qf}{quantum force}

\newcommand{\QM}{quantum mechanics}

\renewcommand{\r}{\rho}
\newcommand{\ra}{\rangle}
\newcommand{\ri}{\rightarrow}

\newcommand{\s}{\sigma}

\newcommand{\se}{Schr\"odinger's equation}

\newcommand{\sv}{slowly varying}
\renewcommand{\t}{\tau}

\newcommand{\vps}{wave packets}
\newcommand{\vp}{wave packet}
\newcommand{\wf}{wave function}

\newcommand{\wl}{wave length}

\newenvironment{deflist}[1]%
{\begin{list}{}%
{\settowidth{\labelwidth}{#1}%
\setlength{\leftmargin}{\labelwidth}%
\addtolength{\leftmargin}{\labelsep}%
\setlength{\parsep}{0pt}%
\setlength{\itemsep}{0pt}%
\setlength{\topsep}{0pt}%
}}%
{\end{list}}%

\begin{document}
\noindent
{\LARGE {\bf Seven Steps Towards the Classical World}}

\vspace{10mm}

\noindent
{\large Valia Allori$^1$,  Detlef D\"{u}rr$^2$, Shelly Goldstein$^3$,
and Nino Zangh\'{\i}$^1$}
\vspace{8mm}

{\footnotesize \begin{deflist}{8}
\item[1]
Dipartimento di Fisica, Istituto Nazionale di Fisica Nucleare - sezione
di Genova,\\ Via Dodecaneso 33, 16146 Genova, Italy\\
e-mail of Allori: allori@ge.infn.it\\
e-mail of Zangh\'{\i}: zanghi@ge.infn.it
\item[2] Mathematisches Institut der Universit\"{a}t M\"{u}nchen,\\
Theresienstra{\ss}e 39,
80333 M\"{u}nchen, Germany\\
e-mail: duerr@rz.mathematik.uni-muenchen.de
\item[3] Departments of Mathematics and Physics,
Rutgers University,\\ New Brunswick, NJ 08903, USA\\
e-mail: oldstein@math.rutgers.edu
\end{deflist}}
\vspace{3mm}
\noindent
November 30, 2001

{\small \paragraph{Abstract.} Classical physics is about real objects,
like apples falling {}from trees, whose motion is governed by
Newtonian laws.  In standard \QM{} only the \wf{} or the results of
measurements exist, and to answer the question of how the classical
world can be part of the quantum world is a rather formidable task.
However, this is not the case for \BM{}, which, like \CM{}, is a theory
about real objects.  In Bohmian terms, the problem of the \cl{}  becomes
very simple: when do the Bohmian trajectories look  Newtonian? } 
\bigskip

\section{Introduction}

The classical world, say the world of objects of familiar experience
that obey Newtonian laws, seems far removed {}from the ``wavy'' world of
quantum mechanics.  In this paper we shall sketch what we believe are
the basic steps  to be taken in going {}from the quantum world
to the classical world.

\paragraph{\rm 1.} The first step is the crucial one: As is well
known, and as Bell has emphasized \cite{bell}, standard \QM{} is not a
precise microscopic theory because the division between the
microscopic and the macroscopic world, which is essential to the very
formulation of that theory, is not made precise by the theory
\cite{goldstein}.  In fact, the following conclusion seems inevitable:
quantum mechanics does not contain the means for describing the
classical world in any approximate sense and one needs to go beyond
\QM{} in order to do so.  There are two natural possibilities for
amending ordinary \QM{}: either the \wf{} is not all there is, or
\se{} is wrong.  In this paper we'll formulate the problem of the
\cl{} within the framework of \BM{}, a theory which follows the first
path and in which the observer doesn't play any crucial role.  It is a
theory about reality, not about the result of measurements.  A very
short review of \BM{} is given in section \ref{sec:bm} and the the
relevant part of this theory, related to the \cl{}, is discussed in
section \ref{sec:clbm}.

\paragraph{\rm 2.} To get a handle on a problem, one should first
simplify it as much as possible.  The complex motion of a
macroscopic body can be drastically simplified by making some rather
standard approximations and reducing the problem to that of a ``particle''
moving in an \extp{}.  This is what we shall do in section
\ref{sec:extp}.

\paragraph{\rm 3.} Good textbooks on \QM{} contain enlightening ideas.
One of these ideas is the so called Ehrenfest theorem that we shall use in
section \ref{sec:wps} in order to obtain a necessary condition for the
classicality for wave packets.

\paragraph{\rm 4.} The structure of \BM{} contains the means for extending
the condition for classicality to more general wave functions, namely
wave functions which locally look like plane waves, as we shall see in
section \ref{sec:LPWs}.  In section \ref{sec:LPWsgen} we shall then
show how the problem of classical limit for general \wf{}s can be
reduced to that for local plane waves.

\paragraph{\rm 5.} Simplicity is good, but it has its limitations: the
reduction of the motion of the center of mass to a one body problem
doesn't explain the robustness and stability of classical behavior.
This however can be explained by making the model a little more
realistic, say by including in an effective way the external as well
as the internal environment.  We shall briefly touch this point in
section \ref{sec:env}.

\paragraph{\rm 6.} This step is the crucial one {}from a mathematical
point of view.  In section \ref{sec:conj} we shall put forward a
mathematical \conj{} on the emergence of classical behavior. 
Unfortunately we cannot provide any rigorous mathematical
justification for it.  Mathematical work on it would be valuable since
this conjecture goes beyond the standard mathematical work of
semiclassical analysis (see, e.g., \cite{maslov}, \cite{robert}) or,
in more modern terms, microlocal analysis (see, e.g.,
\cite{martinez}).

\paragraph{\rm 7.} This is the last step in what we believe is the main
structure of the \cl{}:
$$\left(\psi, X\right)\ri \left(P,X\right),$$
where on the two sides of the arrow are represented the complete state
description of \BM{}, in terms of \wf{} and position, and of \c{}
mechanics, in terms of momentum and position.

\section{Bohmian Mechanics}
\label{sec:bm}

In nonrelativistic \BM{} the world is described by particles which
follow trajectories determined by a law of motion.  The evolution of
the positions of these particles is guided by the \wf{} which itself
evolves according to \se{}.  In other words, in \BM{} the complete
description of the state of an $N$-particle system is the pair
$(\Psi,Q)$, where $\Psi=\Psi(q)=\Psi(q_1,\dots,q_N)$ and $Q=(
Q_1,...,Q_N)$ are respectively the \wf{} and the \emph{actual}
configuration of the system, with $Q_k$ denoting the position of the
$k$-th particle in ordinary three-dimensional space.

For non relativistic spinless particles the state $(\Psi,Q)$ evolves
according to the equations
\begin{eqnarray}
      \ode{Q_k}{t} &  = &\frac{\h}{m_k}
\Im \frac{\nabla_{q_k} \Psi (Q)}{\Psi (Q)}
    \label{eq:bohmeq1}  \\
    i\hbar\pde{\Psi}{t} & = &
    -\sum_{k=1}^N\frac{\h^2}{2m_k}\nabla_{q_k}^2\Psi+ U(q)\Psi
     \label{eq:scevol1}
\end{eqnarray}
Equations (\ref{eq:bohmeq1}) and (\ref{eq:scevol1}) form a complete
specification of the theory.  Agreement between \BM{} and \QM{}
regarding the results of any experiment is guaranteed by what has been
called \cite{qe} the {\it {\q{} equilibrium hypothesis}}: when a
system has a \wf{} $\psi$, its configuration $Q$ is random with \pd{}
\be 
\r(q) =|\psi(q)|^2.
\label{eq:equivmeas}
\ee While the meaning and justification of this hypothesis is a
delicate matter, which has been discussed at length elsewhere
\cite{qe} (see also \cite{durr} and \cite{cushing}), we wish to
underline here an important property of (\ref{eq:equivmeas}): if the
probability density for the configuration satisfies
$\r(q,t_0)=|\psi(q,t_0)|^2$ at some time $t_0$, then the density at
any time $t$ to which this is carried by the motion (\ref{eq:bohmeq1})
is also given by $\r(q,t)=|\psi(q,t)|^2$.  This is an extremely
important property of any Bohmian system, expressing a compatibility
between the two equations of motion (\ref{eq:bohmeq1}) and
(\ref{eq:scevol1}) defining the dynamics, which we call the {\it
equivariance} of $|\psi|^2$.

\section{Motion in an External Potential}
\label{sec:extp}

Our goal is to study the classical behavior of a  macroscopic body
composed of $N$ particles with $N\gg 1$ (one may think of an apple
falling {}from a tree or a planet moving around the sun).  It is rather
clear that one expects classical behavior only for appropriate
macroscopic functions of the particle configuration $(Q_1,...,Q_N)$.
The relevant macroscopic variable, whose classical behavior we wish to
investigate here, is the center of mass of the body
$$X=\frac{\sum_i m_i Q_i}{m}\,,$$
where $m_1,\ldots, m_N$ are the masses of the particles composing the
body and $m = \sum_i m_i$ is the total mass of the body.

We shall assume that the particles interact through internal forces as
well as being subjected to an external potential, so that the
potential energy in (\ref{eq:scevol1}) is of the form
$$ U(q) = \sum_{i<j} U(q_i,q_j) + \sum_i V_i(q_i)\,.  $$
Let $y=(y_1,\ldots,y_{N-1})$ be a suitable set of
coordinates\footnote{For sake of concreteness one may think, e.g., of
the so called Jacobi coordinates.} relative to the center of mass $x=
\sum_i m_i q_i/m$.  Then under the change of variables $q=(x,y)$ \se{}
(\ref{eq:scevol1}) assumes the form \be i\hbar\pde{\Psi}{t} =
\left(H^{x}+ H^{y} +H^{(x,y)}\right) \Psi
\label{eq:sss}
\ee
where
$$
H^x=\frac{\hbar^2}{2m}{\nabla_x}^2 +V(x)\,, \qquad V(x)\equiv \sum_i
V_i(x)\,,
$$
$H^{y}$ is the free Hamiltonian associated with the relative
coordinates $y$ and the operator $H^{(x,y)}$ describes the interaction
between the \com{} and the relative coordinates.  If $V_i$ are \sv{}
on the size of the body, $ H^{(x,y)}$ can be treated as a small
perturbation, and, in first approximation, neglected.  Thus, if
$\Psi=\psi(x)\phi(y)$ at some time, the time evolution of the center
of mass decouples {}from that of the relative coordinates and we end
up with a very simple one particle problem: the \wf{} $\psi$ of the
center of mass evolves according to one-particle \se{} 
\be
i\hbar\pde{\psi}{t}=\frac{\hbar^2}{2m}{\nabla_x}^2\psi +V(x)\psi
\label{eq:scevol}
\ee
and its position $X$ evolves according to
\be
\frac{dX}{dt}=\frac{\h}{m}{\rm Im} \frac{\nabla_x \psi(X)}{\psi(X)}\,.
\label{eq:bohmeq}
\ee
{}From now on, whenever no ambiguity will arise, we shall treat the
center of mass as a ``particle'' and we shall refer to $X$ and $\psi$
as the position and the wave function of such a particle.

\section{The Classical Limit in Bohmian Mechanics}
\label{sec:clbm}

In order to investigate the conditions under which $X$ evolves
classically it is useful to write the \wf{} $\psi=\psi(x)$ in the
polar form
\be \psi(x)=R(x)\ex^{\frac{i}{\h}S(x)},
\label{eq:LPW}
\ee {}From \se{} (\ref{eq:scevol}) one obtains, following Bohm
\cite{bohm52}, the continuity equation for $R^2$, 
\be \frac{\partial
R^2}{\partial t}+{\rm{div}}\left[\left( \frac{\nabla_x
S}{m}\right)R^2\right]=0,
\label{eq:qcont}
\ee
and the modified \HJ{} equation for $S$
\be \frac{\partial
S}{\partial t}+\frac{(\nabla_x S)^2}{2m}+
V- \frac{\h^2}{2m}\frac{{\nabla_x}^2 R}{R}=0.
\label{eq:qHJ}
\ee

Note that equation (\ref{eq:qHJ}) is the usual classical \HJ{}
equation with an additional term
\be
V_Q\equiv-\frac{\h^2}{2m}\frac{{\nabla_x}^2 R}{R},
  \label{eq:potential}
\ee
called the quantum potential.  Since $\frac{\nabla_x S}{m}$ is the right
hand side of (\ref{eq:bohmeq}), one then sees that the (size of
the) quantum potential provides a rough measure of the deviation of
Bohmian evolution {}from its classical approximation.

Analogously, consider the modified Newton equation associated with
(\ref{eq:qHJ}), and obtained by differentiating 
both sides of equation (\ref{eq:bohmeq}) with respect to time,
\be 
m\frac{d^{2}X}{d\,t^{2}}= F +F_Q,
\label{eq:newtonequation}
\ee
where $F=-\nabla_x V(X)$ and $F_Q =-\nabla_x V_Q(X)$ are respectively
the \emph{classical} force and the ``\emph{quantum}'' force.  Equation
(\ref{eq:newtonequation}) shows that all the {\it {deviations}} {}from
classicality are embodied in the \qf{} $F_Q$.

Thus, the \emph{formulation} of the classical limit in \BM{} turns out to
be rather simple: classical behavior emerges whenever the particle
trajectory $X=X(t)$, satisfying (\ref{eq:newtonequation}), approximately
satisfies the classical Newton equation, i.e., \be m\frac{d^{2}X}{d\,t^{2}}
\simeq F \,.
\label{eq:newtoneq}\label{eq:concl}
\ee
The problem is to determine the physical conditions ensuring
(\ref{eq:concl}).  Usually, physicists consider classical behavior as
ensured by the limit $\h\ri 0$, meaning by this \be \h\ll A_0,
\label{eq:ha0} \ee where $A_0$ is {\it some} characteristic action of
the corresponding \cm{} (see, e.g.,
\cite{maslov},\cite{schiff},\cite{berry})).  Condition (\ref{eq:ha0})
is often regarded as equivalent to another standard condition of
classicality which involves the length scales of the motion (see,
e.g., \cite{landau}): if the \db{} \wl{} $\l$ is small with respect to
the characteristic dimension $L$ determined by the scale of variation
of the potential $V$, the behavior of the system should be close to
the \cb{} in the same potential $V$.  This is very reminiscent of how
geometrical optics can be deduced {}from wave optics.  We regard this
condition, i.e.,
\begin{equation}
     \l \ll L,
     \label{eq:lll}
\end{equation}
as the most natural condition of classicality since it relates in a
completely transparent way a property of the state, namely its \db{}
\wl{} $\l$, and a property of the dynamics, namely the scale of
variation of the potential $L$.  In the remainder of this paper we
shall argue that (\ref{eq:lll}) is indeed a necessary and sufficient
condition for (\ref{eq:concl}).

\section{Wave Packets}
\label{sec:wps}

To explain the physical content of (\ref{eq:lll}) and its implications
we shall consider first the case for which the wave function has a
well-defined \db{} \wl{}: we shall assume that $\psi$ is a wave packet
with diameter $\sigma$, with mean wave vector $k$ and
associated  \wl{} $\l =2\pi/|k|$.

As we shall see, the analysis of this situation will allow us to find
a precise characterization of the scale $L$ of variation of the
potential.  Our analysis will be rather standard---it is basically the
\emph{Ehrenfest's Theorem}---and can be found in good textbooks (see,
e.g., \cite{gottfried}).  We reproduce it here both for the sake of
completeness and because we believe that it attains, within the
Bohmian framework, a deeper and much more general significance than
within standard formulations of quantum mechanics.

{}From the equivariance of (\ref{eq:equivmeas}) we have that the mean
particle position at time $t$ is given by
$$
\la X\ra = \int x |\psi_t(x)|^2 dx\,.
$$
{}From (\ref{eq:scevol}) it follows that
$$ m \frac{d^2}{dt^2}\la{}X\ra
= - \int \nabla_x V(x) |\psi_t(x)|^2 dx\,.
$$
By expanding $F(x)=- \nabla_x V(x)$ in Taylor series around $\la X\ra$
one obtains
\be m\frac{d^2 }{dt^2} \la X\ra =F(\la X\ra) + \frac{1}{2}
\sum_{j,k} \Delta_{j,k} \frac{\partial^2 F}{\partial x_j\partial x_k}
(\la X\ra) + ...  ,
\label{eq:taylornewton}
\ee
where $$\Delta_{j,k}=\la X_j X_k \ra - \la X_j \ra \la X_k \ra $$
is of order $\sigma^2$, where $\sigma$ is the diameter of the packet.
Therefore, the mean particle position should satisfy the classical Newton
equation whenever
\be \s^2 \left| \frac{\partial^3 V}{\partial
x_i\partial x_j\partial x_k}\right| \ll \left|\frac{\partial
V}{\partial x_i } \right| \,,
\label{eq:condition}
\ee
i.e.,
\be
\s\ll\sqrt{\left|\frac{V'}{V'''}\right|}
\label{eq:cond1}
\ee
where $V'$ and $V'''$ denote respectively suitable estimates of
the first and third derivatives (e.g., by taking a sup over the
partial derivatives).

The minimum value of the diameter of the packet $\sigma$ is of order
$\l$.  Hence (\ref{eq:cond1}) becomes
\be
\l\ll\sqrt{\left|\frac{V'}{V'''}\right|}
\label{eq:cond2}
\ee
This last equation gives a \emph{necessary} condition for the 
classicality of the particle motion and, by comparing it with
(\ref{eq:lll}), a precise definition of the notion of scale of
variation of the potential, namely,
\be L= L(V)
=\sqrt{\left|\frac{V'}{V'''}\right|}\,.
\label{eq:L}
\ee
\medskip

In the following we shall argue that (\ref{eq:lll}), with $L$ given by
(\ref{eq:L}), is indeed also {\em sufficient} for classical behavior
of Bohmian trajectories.  For wave packets this follows easily {}from
the equivariance of $|\psi|^2$: over the lapse of time for which the
spreading of the packet can be neglected, the overwhelming
majority\footnote{With respect to the equivariant measure $|\psi|^2$.}
of trajectories $X=X(t)$ will stick around their mean value $\la X\ra$
and follow its classical time evolution.  Thus we expect
(\ref{eq:concl}) to hold for the overwhelming majority of
trajectories.

\section{Local Plane Waves}
\label{sec:LPWs}

Suppose now that $\psi$ is not a packet but a wave function that
locally looks like a packet.  By this we mean, referring to the polar
representation (\ref{eq:LPW}), that the amplitude $R(x)$ and the {\em
local} wave vector 
\be 
k=k(x)\equiv \nabla_x S(x)/\hbar
\label{eq:lwl}
\ee 
are \sv{} over distances of order $\l(x)\equiv{h}/{|\nabla_x
S(x)|}$, the {\em local} de Broglie \wl{}.  We may call such a $\psi$
a ``local plane wave''.

At any given time the \LPW{} can be thought as composed of a sum of
\vps{}: Consider a partition of physical space into a union of
disjoint sets $\D_i$ chosen in such a way that the local wave vector $k(x)$
doesn't vary appreciably inside of each of them and denote by $k_i$
the almost constant value $k(x)$ for $x\in \D_i$.  Let $\chi_{\D_i}$
be the characteristic function of the set $\D_i$ ($\chi_{\D_i}(x)=1$ if
$x\in \D_i$ and $0$ otherwise).  Since $\sum_i \chi_{\D_i}=1$, we have
\be
\psi(x)=\sum_i \chi_{\D_i}(x)\psi(x)=\sum_i \psi_i (x).
\label{eq:sum}
\ee 
Note that this decomposition is somewhat arbitrary: provided that
$k(x)$ is almost constant in $\D_i$, the extent of these sets can be
of the order of many wave lengths down to a minimal size
$\s_i\simeq|\Delta_i|^{1/3}$ of the same order of $\l_i$.\footnote{The
use of the characteristic function may introduce an undesirable lack
of smoothness, but this can be easily taken care by replacing the
$\chi_{\D_i}$ with functions $\theta_i$ forming a smooth partition of
unity.}

At any time, the position $X$ of the particle will be in the support
of one of the packets forming the decomposition (\ref{eq:sum}), say in
the support of $\psi_i$.  If the condition (\ref{eq:cond1}) holds for
$\s_i$, we may then proceed as in the previous section: the minimal
size of the packet $\psi_i$ can be taken of order $\s_i=\l(x)$ and the
condition of classicality is again (\ref{eq:cond2}) for $\l=\l(x)$.

Note that this straightforward reduction of the classical limit for
local plane waves to that for wave packets is possible only within
\BM{}: since the particle has at any time a well-defined position $X$
and the different components of the local plane wave (\ref{eq:sum})
don't interfere, we may ``collapse'' $\psi$ to the wave packet
$\psi_i$ relevant to the dynamics of $X$.

\section{General Wave Functions}
\label{sec:LPWsgen}

We wish now to investigate the physical content of (\ref{eq:lll}) and
its implications for a general \wf{}.  The first issue to address is
what notion of \wl{} should be appropriate for this case.  A rough
estimate of $\l$ could be given in terms of mean kinetic energy
associated with $\psi$, 
\be 
E_{\rm kin} (\psi)= \la\psi,
-\frac{\h^2}{2m}{\nabla_x}^2\psi\ra\,,
\label{kinen}
\ee
with associated wave length
\be
\l =\l(\psi) =\frac{h}{ \sqrt{ 2m E_{\rm kin}(\psi) } }\,.
\label{eq:lambda}
\ee

Suppose now that (\ref{eq:cond2}), with $\l$ given by
(\ref{eq:lambda}), is satisfied.  We claim that in this case the
Schr\"odinger evolution should ``quickly'' produce a local plane wave,
that can be effectively regarded as built of pieces that are wave
packets satisfying (\ref{eq:cond2}) for $\l=\l(x)$ and hence
themselves evolving classically as we have seen in the previous
section.

In fact, if $\l\ll L$ the kinetic energy dominates the potential
energy and the free Schr\"odinger evolution provides a rough
approximation of the dynamics up to the time needed for the potential
to affect the evolution significantly.  During this time, the
Schr\"odinger evolution produces a spatial separation of the different
wave vectors contained in $\psi$, more or less as Newton's prism
separates white light into the different colors of the rainbow.  In
other words, the formation of a local plane wave originates in the
``{\em dispersive}'' character of free Schr\"odinger evolution.

So, in order to gain some appreciation of this phenomenon consider the
free Schr\"odinger evolution
\begin{equation}
   \psi_t(x)=\frac{1}{(2\pi)^{3/2}}\int
   \ex^{it\left[k\frac{x}{t}-\frac{\h{k}^2}{2m}\right]}
   \hat{\psi}(k)dk\,,
  \label{eq:psixt}
\end{equation}
where $\hat{\psi}$ is the Fourier transform
of the initial \wf{} $\psi$.  The stationary phase method yields
straightforwardly the long time asymptotics of $\psi_t$,
\begin{equation}
  \psi_t(x)\sim  \left(\frac{im}{\h t}\right)^{3/2}
   \ex^{i\frac{m}{2\h}\frac{{x}^2}{t}}
   \hat{\psi}(k)\,,\quad {\rm where}\quad k=\frac{m}{\h}\frac{x}{t}\ ,
  \label{eq:spm}
\end{equation}
which is indeed a local plane wave with local wave vector 
$k=mx/(\hbar t)$.

We said above that the local plane wave is ``quickly'' produced.  But
how quickly?  In order to estimate such a time, consider the
simple example of an \iwf{} $\psi$ composed of two overlapping \vps{}
with the same position spread $\Delta x$ and with opposite momenta $p$
and $-p$.  The time $\tau$ of formation of a local plane should be of
the order of the time for separation of the packets, which is basically
the time needed to cover a space equal to $\Delta x$.  {}From $\Delta
x \Delta p\sim\h$ and $\Delta p\sim p$ we obtain
\be \t \sim
\frac{\Delta x}{p/m} \sim \frac{\h}{p^2/m}\sim \frac{\h}{\la E\ra},
\label{eq:timelpw}
\ee
where $\la E\ra $ is the mean kinetic energy of the particle.  It
is reasonable to suggest that (\ref{eq:timelpw}), with $\la E\ra $
given by (\ref{kinen}), could give a very rough estimate of the time
of formation of a local plane wave for a general \wf{} $\psi$.  Note
that the time needed for the potential to produce significant effects
on the evolution is of order \be T=\frac{L}{v},\qquad {\rm where}
\qquad v=\frac{h}{m\l}.
\label{eq:T}
\ee
Thus, if $\l\ll L$ we have that $ \t\ll T$, which means that the
local plane wave gets formed on a time scale much shorter than the
time scale over which the potential affects the dynamics.  
\medskip

We arrive in this way at a sharp (or, at least, sharper than usually
encountered) mathematical formulation of the classical limit for a
general wave function $\psi$.  First of all, consider the
dimensionless parameter 
\be 
\e=\frac{\l({\psi})}{L(V)}\,.
\label{eq:epsilon}
\ee
Secondly, consider the Bohm motion $X$ on the ``macroscopic" length
and time scales defined by $\psi$ and $V$.  By this we mean
$X'=X'(t'),$ where
\be X' = X/ L\quad {\rm and}\quad t' = t/T
\label{eq:macroscales}
\ee
with $T$ given by (\ref{eq:T}).  Finally, consider $ F_Q/m $,
the ``quantum'' contribution to the total acceleration in
(\ref{eq:newtonequation}), on the macroscopic scales
(\ref{eq:macroscales}), namely 
\be 
 D =\frac{T^2}{L} F_Q (X' L, t'T)
\label{eq:deviation}
\ee
Then the Bohm motion on the macroscopic length and time scales
will be approximately classical, with deviation {}from classicality $D$
tending to $0$ as $\e \to 0$.

We'd like to point out that the use of macroscopic coordinates
(\ref{eq:macroscales}) for the formulation of the classical limit is
rather natural {}from a physical point of view.  First of all, the
scales $L$ and $T$ are the fundamental units of measure for the
motion: $L$ is the scale on which the potential varies and $T$
provides an estimate of the time necessary for the particle to see its
effects.  More importantly, in the limit $\e\to 0$ the nonclassical
behavior---occurring during the time $\tau$ of formation of the local
plane wave---disappears, since, as we have argued above, in this limit
$\tau\ll T$.  In other words, on the macroscopic scales on which we
expect classical behavior the local plane wave has been formed.

\section{Limitations of the Model: Interference and the Role of the Environment}
\label{sec:env}

Before commenting on the mathematics of the limit $\e\to0$ we should
stress a physical caveat.  For motion in unbounded space, the
expanding character of the Schr\"odinger evolution makes the set of
local plane waves an ``attractor'' for the dynamics---so that the
\LPW{} form is in this sense ``typical''.  However, for motion in a
bounded region (with \wf{}s which are superpositions of bound states)
the ``typical'' \wf{} is composed by a sum of \LPWs{}, this being due
to interference between the waves reflected by the ``edges'' of the
confining potential.  Consider for example an infinite potential well
of size $L$ in one dimension and initial \wf{} $\psi$, well localized
in the center of the well which is the superposition of two packets
with opposite momenta $p$ and $-p$.  Suppose that $\l(\psi)\ll L$. 
Then the two packets move classically and at a certain time, say
$t_{r}$, are reflected {}from the walls of the potential.  At the time
$t_c=2t_{r}$, they interfere in the middle of the well.  $t_c$ is the
``{\em first caustic time},'' the time at which the classical action
$S_{\rm cl} (x,t)$ becomes multivalued.  In general, we should not
expect classical behavior for times larger than the first caustic time
$t_c$.

What is going on?  The emergence of classical behavior should be
robust and stable, which would not be the case if it were restricted
to times smaller than $t_c$.  However, if one remembers that the model
we are investigating is a strong idealization, the problem evaporates. 
We are in fact dealing with the one-body problem defined by
(\ref{eq:scevol}) and (\ref{eq:bohmeq}), an approximation to the
complete dynamics defined by (\ref{eq:sss}) in which the term
$H^{(x,y)}$, describing the interaction between the \com{} and the
relative coordinates, is neglected.  Note than even (\ref{eq:sss}) is
an idealization since it does not include the unavoidable interaction
of the body with its {\em external} \env{}: in a more realistic model
$H^{(x,y)}$ would take into account both the internal and external
environment of the center of mass (with $y$ now including both the
relative coordinates and the degrees of freedom of the external
\env{}).  These interactions---even for very small interaction
energy---should produce {\it entanglement} between the \com{} $x$ of
the system and the other degrees of freedom $y$, so that their
effective role is that of ``measuring'' the position $X$ and
suppressing superpositions of spatially separated \wf{}s.  (Taking
these interactions into account is what people nowadays call \deco{},
see, e.g., \cite{libro} and the references therein).  Referring to the
above example, the effect of the environment should be to select (as
relevant to the dynamics of $X,$ see \cite{qe} and \cite{shelly}) one
of the two packets on a time scale much shorter than the first caustic
time $t_c$.

\section{Towards a Mathematical Conjecture}
\label{sec:conj}

The mathematical content of sections \ref{sec:LPWsgen} and
\ref{sec:env} is summarized by the following (not yet sharply
formulated) conjecture: 
\medskip

\noindent\textbf{Conjecture.} {\em Let $\e$ be the dimensionless
parameter defined by (\ref{eq:epsilon}) and $D$ be the quantity given
by (\ref{eq:deviation}).  Then there are environmental interactions
such that $D\leadsto 0$ as $\e\to 0$, uniformly in $\psi$ and $V$.}
\bigskip

Concerning this conjecture, we'd like to make here just a few remarks.

\paragraph{\rm 1.} ``$D\leadsto 0$'' means convergence to 0 in a
``suitable'' probabilistic sense since $D$ is a random variable.  $D$
is a function of $X$, and $X$ is random with probability distribution
given by $|\psi|^2$.  To require almost sure convergence is probably
too strong a demand.  Convergence in probability, or $L^2$
convergence, would seem more appropriate.  Moreover, $\l$ is defined
in (\ref{eq:lambda}) in terms of the {\it average\/} kinetic energy. 
This average could be large even when there is a significant
probability for a very small kinetic energy.  Thus it is probably
necessary to regard $\l$ as random (with randomness inherited {}from
the kinetic energy) and to understand $\e\to 0$ also in a
probabilistic sense.

\paragraph{\rm 2.} Uniformity of the limit in $\psi$ and $V$ could be
expressed as follows: let $(V_n,\psi_n)$ be any sequence for which
$\e_n=\frac{\l_n}{L_n}\ri 0$, with $\l_n=\l(\psi_n)$ given by
(\ref{eq:lambda}), and $L_n=L(V_n)$.  Then $D\leadsto 0$ as $n\ri
+\infty$.  Understanding $D\leadsto 0$ as convergence in probability,
we could also express uniformity in the following way: for any
$\eta>0$ and for any $\d>0$, there exists an $\e_0>0$ such that
$\mathbf{P}(D>\d)$ is smaller than $\eta$ whenever $\e<\e_0$.  Here
$\textbf{P}$ is the probability measure defined by $|\Psi|^2$, i.e.,
$\textbf{P}(dq)= |\Psi(q)|^2dq$, which includes randomness arising
{}from the environment.

\paragraph{\rm 3.} For quadratic potentials (including free motion and
motion in a uniform force field) $L=\infty$ so that $\e=0$.  In this
case the \conj{} should be modified as follows: let $L_o$ be {\it any}
length scale and $T_o$ the corresponding time scale
$T_o=\frac{L_o}{v}$.  Then for the Bohm motion on the scales given by
$L_o$ and $T_o$, $D\leadsto 0$ uniformly in $\psi$ and $L_o$ whenever
$\tilde{\e}\equiv\frac{\l(\psi)}{{L_o}}\ri 0$.

\paragraph{\rm 4.} There is an enormous amount of mathematical work,
called semiclassical analysis or, in more modern terms, microlocal
analysis, in which the limit $\hbar\to 0$ of Schr\"odinger evolutions
is rigorously studied.  It should be stressed that the limit $\e\ri 0$
is {\em much more general} than the limit $\h\ri 0$.  In fact
$\e=\l/L=h/mvL$.  So keeping $L$ and the momentum $mv$ fixed, the
limit $\h\ri 0$ implies $\e\ri 0$.  But there are many ways in which
$\e$ could go to zero.  The \cl{}, as expressed by the above
conjecture, is (at the very least) a two-parameters limit, involving
$\l$ and $L$, and $\h\ri 0$ is just a very special case.  Moreover,
these two parameters themselves live on infinite dimensional spaces
since $\l=\l (\psi)$, with $\psi$ varying in the Hilbert space of the
system's wave functions, and $L=L (V) $, with $V$ varying in the class
of admissible one particle potentials (that is, potentials leading to
a self-adjoint Hamiltonian).

\paragraph{\rm 5.} Exactly for the reason expressed in the previous
remark, the \conj{} is really {\em very} hard to prove: it require a
lot of uniformity both in the \wf{} $\psi$ and in the potential $V$. 
Just to have an idea of the difficulties, one may think of the
analogous problem in statistical mechanics, namely the problem of
studying the deviations {}from thermodynamic behavior of a large but
finite system about which not so much is known.

\paragraph{\rm 6.} While the \conj{} is difficult to prove, it is
still not completely satisfactory {}from a physical point of view. 
The \conj{} states only that $D$ depends on $\e$ in such a way that
$D\leadsto 0$ as $\e\ri 0$, uniformly in $\psi$ and $V$.  A physically
more relevant result would be to estimate how rapidly $D$ is tending
to 0 (e.g., like $\e$, or $\e^2$ or whatever).  Note that only this
last kind of result can be of practical value: given $V$ and $\psi$,
it provides an estimate for the \dfc{}, while any other results do not
quite do this.

\paragraph{\rm 7.} With the conjecture, and even with the refinement
proposed in the previous remark, there is a further difficulty to
consider: even if $H^{(x,y)}$ is treated as a small perturbation in
(\ref{eq:sss}), the suggestion of section \ref{sec:env} might not be
too realistic.  In fact, the autonomous Schr\"odinger evolution, even
of a very narrow \wf{} $\psi=\psi(x)$, could be destroyed in very
short times.  This is a serious difficulty; one resolution might be
found in the notion of {\em conditional wave function} of the
$x$-system, $\psi(x) = \Psi(x, Y)$, where $Y$ is the actual
configuration of the environment (this notion has been introduced and
analyzed in \cite{qe}).  We regard the extension of the conjecture to
this more realistic framework as the most interesting open problem on
the classical limit, which we leave for future work.

\section{The Classical Limit in a Nutshell}
\label{sec:nut}

The key ingredient in our analysis of the emergence of the \c{} world,
is that as soon as the \LPW{} has formed, each configuration $X$ is
attached to a guiding \vp{} with a definite wave vector $k(x,t)$ that
locally determines the particle dynamics according to the local de
Broglie relation
$$p(x,t)=\hbar k(x,t),$$
which, for $\l\ll L$, evolves according to \c{} laws.  This means that
the \cl{} can be symbolically expressed as
$$(\psi,X)\ri (P,X), $$
where $(\psi,X)$ is the complete quantum state description in terms of
\wf{} and position, while $(P,X)$ is the complete \c{} state
description in terms of momentum and position.  All the relevant
macroscopic information contained in the pair $(\psi,X)$ is, in the
classical limit, embodied in the pair $(P,X)$---the only robust,
stable quantity.  In other words, as far as the macroscopic dynamics
of $X$ is concerned, only the information carried by $P$ is relevant.

\section*{Acknowledgments}

This work was financially supported in part by the INFN and the DFG.
Part of the work has grown and has been developed in the Dipartimento
di Fisica dell'Universit\`a di Genova, the IHES of Bures sur Yvette,
the Mathematisches Institut der Universit\"at M\"unchen and the
Department of Mathematics of Rutgers University.  The hospitality of
these institutions is gratefully acknowledged.  We thank Herbert Spohn
and Roderich Tumulka for helpful discussions.  Finally, we thank
Stefan Teufel and James Taylor for a continuous exchange of ideas and
their involvement in a larger, common project on the derivation of the
classical limit.

\end{document}